Research Article

Xiao Guo, Xin He, Zachary Degnan, Chun-Ching Chiu, Bogdan C. Donose, Karl Bertling, Arkady Fedorov, Aleksandar D. Rakić* and Peter Jacobson*

# Terahertz nanospectroscopy of plasmon polaritons for the evaluation of doping in quantum devices

**Abstract:** Terahertz (THz) waves are a highly sensitive probe of free carrier concentrations in semiconducting materials. However, most experiments operate in the far-field, which precludes the observation of nanoscale features that affect the material response. Here, we demonstrate the use of nanoscale THz plasmon polaritons as an indicator of surface quality in prototypical quantum devices properties. Using THz near-field hyperspectral measurements, we observe polaritonic features in doped silicon near a metal-semiconductor interface. The presence of the THz surface plasmon polariton indicates the existence of a thin film doped layer on the device. Using a multilayer extraction procedure utilising vector calibration, we quantitatively probe the doped surface layer and determine its thickness and complex permittivity. The recovered multilayer characteristics match the dielectric conditions necessary to support the THz surface plasmon polariton. Applying these findings to superconducting resonators, we show that etching of this doped layer leads to an increase of the quality factor as determined by cryogenic measurements. This study demonstrates that THz scattering-type scanning near-field optical microscopy (s-SNOM) is a promising diagnostic tool for characterization of surface dielectric properties of quantum devices.

# 1 Introduction

To achieve a quantum advantage over classical computers, the fabrication of high-fidelity qubits with long dephasing time and low coherence loss is required [1–3]. While superconducting qubits are one of the most mature quantum computing platforms, the realisation of their full potential is currently limited by material losses, many of which are introduced during fabrication. In practice, the primary decoherence mechanisms are concentrated on lossy dielectrics at device interfaces, particularly the substrate-air and metal-air interfaces. These dielectric losses limit the low temperature performance of superconducting devices, stunting the progress of scalable quantum computing [3, 4]. Silicon is the industry-standard substrate for superconducting and spin qubits [5, 6]. However, to quantify and minimize losses introduced during fabrication, non-destructive diagnostic methods are needed for the nanoscale characterization of quantum relevant surfaces and interfaces.

Surface plasmon polaritons (SPPs) are a type of quasi-particles formed by the collective coupling of incident electromagnetic waves to free electrons and are bound to dielectric conductor interfaces [7, 8]. Due to their strong confinement at interfaces, SPPs form the basis of high-precision sensors that can detect minute changes in the local

*Corresponding authors: Aleksandar D. Rakić, School of Information Technology and Electrical Engineering, The University of Queensland, St Lucia, Brisbane, 4072, QLD, Australia, E-mail: a.rakic@uq.edu.au.
https://orcid.org/0000-0002-4615-2240; and **Peter Jacobson**, School of Mathematics and Physics, The University of Queensland, St Lucia, Brisbane, 4072, QLD, Australia, E-mail: p.jacobson@uq.edu.au.
https://orcid.org/0000-0002-5363-3763
**Xiao Guo**, **Bogdan C. Donose** and **Karl Bertling**, School of Information Technology and Electrical Engineering, The University of Queensland, St Lucia, Brisbane, 4072, QLD, Australia,
E-mail: xiao.guo@uq.edu.au (X. Guo), b.donose@uq.edu.au (B.C. Donose), k.bertling@uq.edu.au (K. Bertling). https://orcid.org/0000-0003-4864-0472 (X. Guo). https://orcid.org/0000-0003-1935-9489 (K. Bertling)
**Xin He**, **Zachary Degnan**, **Chun-Ching Chiu** and **Arkady Fedorov**, School of Mathematics and Physics, The University of Queensland, St Lucia, Brisbane, 4072, QLD, Australia; and ARC Centre of Excellence for Engineered Quantum Systems, St Lucia, Brisbane, 4072, QLD, Australia,
E-mail: x.he@uq.edu.au (X. He), z.degnan@uq.edu.au (Z. Degnan), chunching.chiu@uq.edu.au (C.-C. Chiu), a.fedorov@uq.edu.au (A. Fedorov). https://orcid.org/0000-0002-2913-6388 (X. He). https://orcid.org/0000-0002-3049-6639 (Z. Degnan). https://orcid.org/0000-0003-2005-0390 (C.-C. Chiu). https://orcid.org/0000-0002-9815-0968 (A. Fedorov)

dielectric environment. Therefore, a quantitative approach to evaluate surface dielectric properties via knowledge of surface plasmons is possible. To support SPPs, the real part of the complex permittivity must have the opposite sign across an interface under *p*-polarized electromagnetic radiation [9]. One strategy to meet this condition is doping, which allows control over the polariton dispersion and tune the supported frequency regimes for plasma frequencies within the THz regimes [10]. Therefore, dopants introduced during the fabrication of silicon-based superconducting quantum devices [2, 11] can potentially support of plasmons under THz stimulus. To resolve evanescent light–matter responses, scattering-type scanning near-field optical microscopy (s-SNOM) has emerged as a key nanoscopy tool for long-wavelength radiation, such as THz waves [12–16]. It has been employed to characterize surface-confined electromagnetic responses in microstructures [17], study carriers plasmon and phonon polaritons in van der Waals materials [18–20], and reveal THz polaritons in topological materials [21, 22] at sub-micron resolution.

To begin this study, we report the observation of THz plasmon polaritons in superconducting quantum devices. A multilayer model is considered to describe the fabrication-induced doping layer. To retrieve both the unknown thickness and complex permittivity of the doped layer without the prior assumptions on the form of a permittivity model, a multilayer extraction procedure for THz nanospectroscopy using vector calibration is demonstrated. The observed polaritonic features from THz near-field measurements are due to the fabrication-induced surface doping and thereby due to THz-induced surface plasmon polaritons. Finally, to evaluate the influence of doping on potential loss mechanism for superconducting qubits, low-temperature characterization measurements are performed on the as-prepared and etched quantum devices. Removing the doped layer increases the quality factor of the etched device, demonstrating the utility of the nanoscale characterization method to inform fabrication efforts. Therefore, THz surface plasmon polaritons serve as an indicator of semiconductor device surface quality.

## 2 Results and discussion

### 2.1 THz white-light nanoimaging

Figure 1(a) shows a three-dimensional rendering of the chip under investigation combined with a schematic representation of the s-SNOM geometry above a coplanar microwave resonator in Figure 1(b). The region of interest for our s-SNOM investigations is the etched domain defining the transmission line as shown in Figure 1(c) and this microscopic structure is repeatedly investigated with multiple tips and different samples in different device orientations.

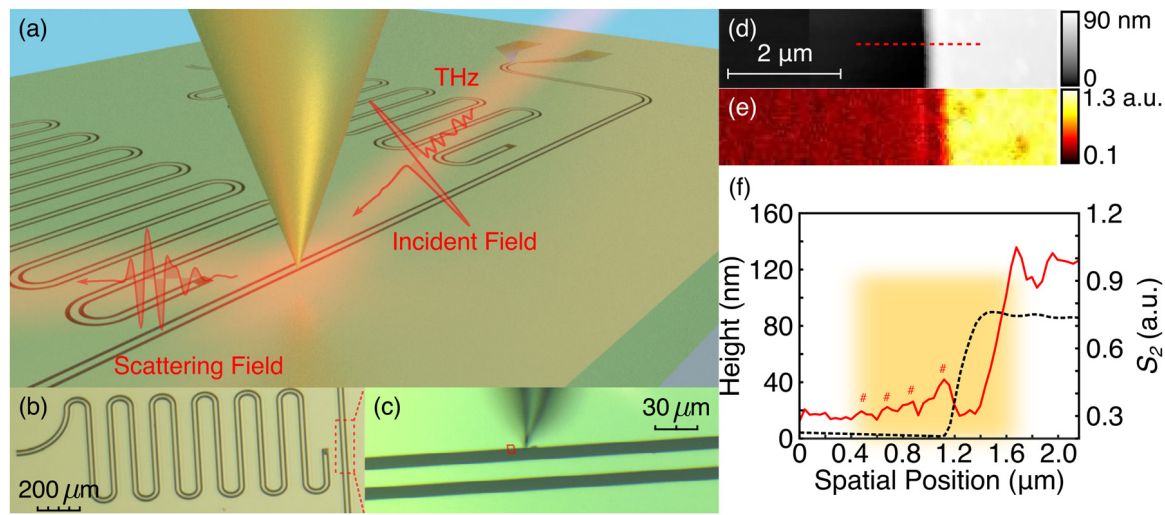

**Figure 1:** THz white-light nanoimaging. (a) Schematic of THz s-SNOM experiments on a chip with coplanar microwave resonators, (b) optical micrograph of an individual coplanar microwave resonator and (c) top view of the area under an optical camera in s-SNOM for further scanning, a scanning area across deposited aluminium film and etched silicon channel: (d) topography, THz near-field scattering signals demodulated at (e) the second-order ($S_2$) harmonics of the probe tapping frequency, and (f) the height profile (black). The corresponding THz white-light scattering amplitudes (red) are obtained from the line scan (45 nm/pixel) denoted by the red broken line in (d) with averaging over 10 adjacent lines. Oscillation peaks are highlighted by markers (# for $S_2$). The orange shaded area in (f) was selected to further perform high-resolution (20 nm/pixel) THz hyperspectral measurements.

Tapping-mode AFM imaging in Figure 1(d) reveals that the device is dominated by a sharp 85-nm step demarcating the aluminium and silicon regions. Turning to the THz scattering signals, we observe a prominent increase in scattered intensity just within the silicon channel for the second-order harmonic ($S_2$) of the signal (Figure 1(f)). By extracting a sequence of line profiles (45 nm/pixel) across the boundary with averaging over 10 adjacent lines (center in the dashed line in Figure 1(d)), we observe a series of oscillations within $S_2$ signals, as shown in Figure 1(f).

## 2.2 THz near-field hyperspectral analysis

To better understand and identify the source of the observed spatial-varying oscillatory behaviour in the channel, we perform a high-resolution (20 nm/pixel) THz hyperspectral scan on a second identical device. This hyperspectral scan crosses the aluminium-silicon interface as shown in Figure 2. This dataset includes both the phase and amplitude information of s-SNOM scattering signals and allows analysis of the imaginary part of reflectivity ($\text{Im}\{r_{exp}\} = \text{Im}\{re^{i\theta}\}$). To reconstruct the complex reflectivity ($r_{exp} = re^{i\theta}$) from the scattering signals ($S_n$), we perform a vector calibration (see Methods). Figure 2(a) shows the spectrally averaged line profile from the calibrated reflectivity, $|r_{exp}|$, between 0.6 and 1 THz (see Figure S3(a)). Similar to the white-light (broadband scattered signals) line profile in Figure 1(g), a series of oscillations is observed well within the silicon channel. The broad transition in the scattered amplitude near the interface is indicative of the tip-induced edge darkening effect, commonly observed in s-SNOM, due to the tip radius (65 nm, from knife-edge tests in Figure S2). To eliminate the spurious feature contribution (edge darkening effect and edge mode) from the tip-edge convolution [23–25], we focus our attention to a region greater than 600 nm away from the interface (highlighted in Figure 2a), in light of the calibrated amplitude and phase contour in Figure S3(b) and (c) to get rid of potential edge-induced phase artefact, to study monochromatic responses on the fringes. Figure 2(b) shows the imaginary part of calibrated reflectivity, Im{$r_{exp}$}, using the spectrally resolved (monochromatic) frequency response (see its contour in Figure S3(d)). The spectrally resolved scattering in this region shows a pronounced sinusoidal oscillation across the sampled spectral range (0.6–1.35 THz). We observe that the fringe spacing varies with the probing frequency and attribute it to the coherent interference between the tip-launched surface waves and the reflection from the interface (Figure 2c).

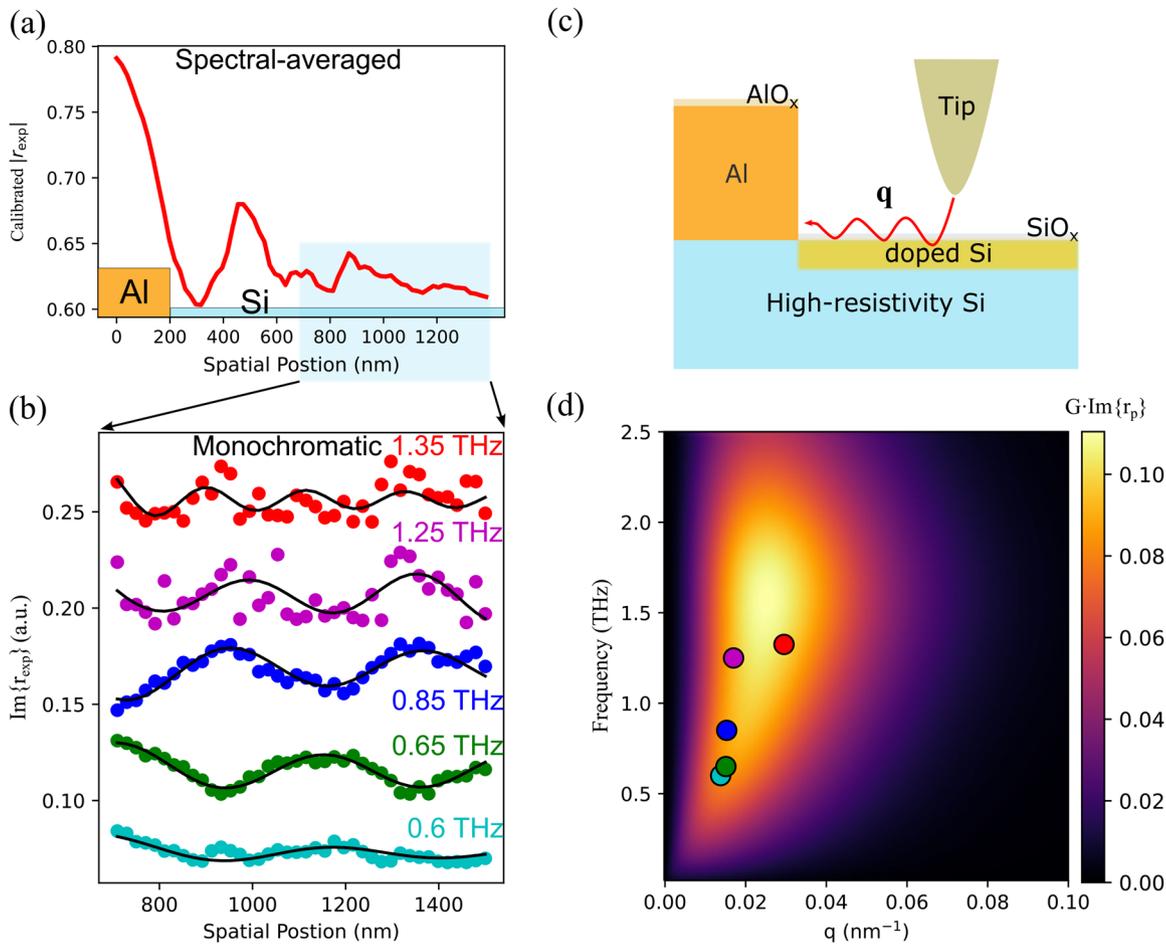

**Figure 2:** THz s-SNOM hyperspectral analysis. (a) A high-resolution (20 nm/pixel) line profile (calibrated reflectivity, $|r_{exp}|$) across the Al-Si interface as shaded in Figure 1(g), (b) near-field line profiles of second-order harmonic near-field scattering signals from high-resolution (20 nm/pixel) hyperspectral measurements of the Si channel (starting from 200 nm away from the Al–Si interface till the far end) showing a dispersive nature (cyan: 0.6 THz, green: 0.65 THz, blue: 0.85 THz, magenta: 1.25 THz, and red: 1.35 THz) of the spatially varying THz tip-sample interactions (vertically offset for clarity), (c) schematic illustration of the nanostructure of the interrogated chip employing THz s-SNOM, (d) the dispersion curve calculated from the imaginary part of p-polarized reflection coefficient considering an Air/SiO$_2$ (2-nm)/doped Si (13-nm)/Si multilayer nanostructure and a typical tip near-field coupling function for s-SNOM tip-sample interactions (see Extended Data, Figure S1). In-plane momenta extracted from fitting profiles (black lines) in (b) are displayed in (d).

To better understand the fringes in the silicon channel, we consider a surface doped layer on a nominally undoped high-resistivity silicon substrate, consistent with earlier measurements of carrier concentration for etched silicon surfaces [11]. Theoretically, the in-plane momentum launched by the s-SNOM probe tip is described by a time-averaged weight function for the duration of the probe's tapping motion, H(t), and follows a bell-shape distribution described by $\langle q^2 \exp(-2qH(t)) \rangle_t$ (see Figure S1) [26]. By focusing an incident THz field on the probe tip, an induced polarization appears on the probe which further polarizes the sample (see such a case in Figure S6(a)). Hence, an iterative polarization process occurs between the probe-sample pair, which can be regarded as a pair of electric dipoles [27] (see an example in Figure S6(c)). The oscillation motion of the tapping probe modulates the distance between this interacting dipole pair and yields re-radiated light accounting for the s-SNOM scattering response. We expect this scattering quantity to be proportional to the scattered probe-sample dipole moment [28], which may be described by finite dipole model (see Figure S6(b) and (d)). Additionally, this quantity contains a frequency dependent enhancement below a certain cut-off frequency originating from the antenna resonance effect in s-SNOM, depending on the probe geometry and the dipolar resonance between the tip-sample pair [29–32]. Therefore, the s-SNOM scattering response is convoluted with the material properties and instrumental responses.

The measured s-SNOM near-field response in our system is a weighted average of the in-plane (p-polarized) momentum-dependent reflection coefficient, $\langle r_p(\omega, q) \rangle_q$.

The coherent-coupling mode in the near-field is characterized by the maxima of Im{$r_p(\omega, q)$} at real-valued $q$ [26]. In the interrogated multilayer system, the combination of thin-film thickness and surface doping effectively alters the plasma frequency to support the excitation of THz polaritons with the tip-launched in-plane momenta. Considering the nanostructure described in Figure 2(c) with frequency-dependent complex dielectric permittivity, its dispersion relation ranging from 0 to 2.5 THz is calculated and shown in Figure 2(d) (details in Methods). By combining the material dependent and tip dependent responses, we can then extract in-plane momenta $q$ that coincide with the observed fringes in our experiment and is consistent with the calculated dispersion relation. These fits are shown in Figure 2(b) (solid lines) and the extracted momenta are shown in Figure 2(d).

## 2.3 Multilayer extraction with THz nanospectroscopy

As the substrate is high-resistivity silicon with a known permittivity at the probing frequency ($\varepsilon_{Si} = 11.6$), the fingerprints of the surface polariton in Figure 2(b) originate from the fabrication-induced surface doping layer reported previously [11]. Hence, a quantitative extraction of the thin-film thickness and permittivity is required to further validate the surface dielectric condition to support THz polaritons. We developed a multilayer extraction procedure which follows our previous work describing calibration of SNOM for direct permittivity measurements [33]. The details of this procedure are explained in detail below.

To quantitatively extract both the thickness and complex permittivity of the doping layer, we extend our previous s-SNOM calibration method to the characterization of multilayers. This process is described schematically in the flow diagram shown in Figure 3.

The steps to follow for the characterization of a multilayer structure using an s-SNOM are summarised in Figure 3(a):

- (1) obtain $n$th order harmonic scattering spectra ($S_n$) of the interrogated sample (red: amplitude, blue: phase) from s-SNOM.
- (2) calibrate s-SNOM system responses ($e_n$) for $n$th order harmonic signals. Use the obtained three error terms ($e_D, e_R, e_S$) to calibrate the interrogated sample response ($S_n$) and to obtain the calibrated reflection coefficient ($\beta_{exp}$, red: amplitude, blue: phase) as the averaged response of a multilayer structure.
- (3) decouple the contribution of a thin film ($\varepsilon_2$, $d$) from the calibrated reflection coefficient with the prior

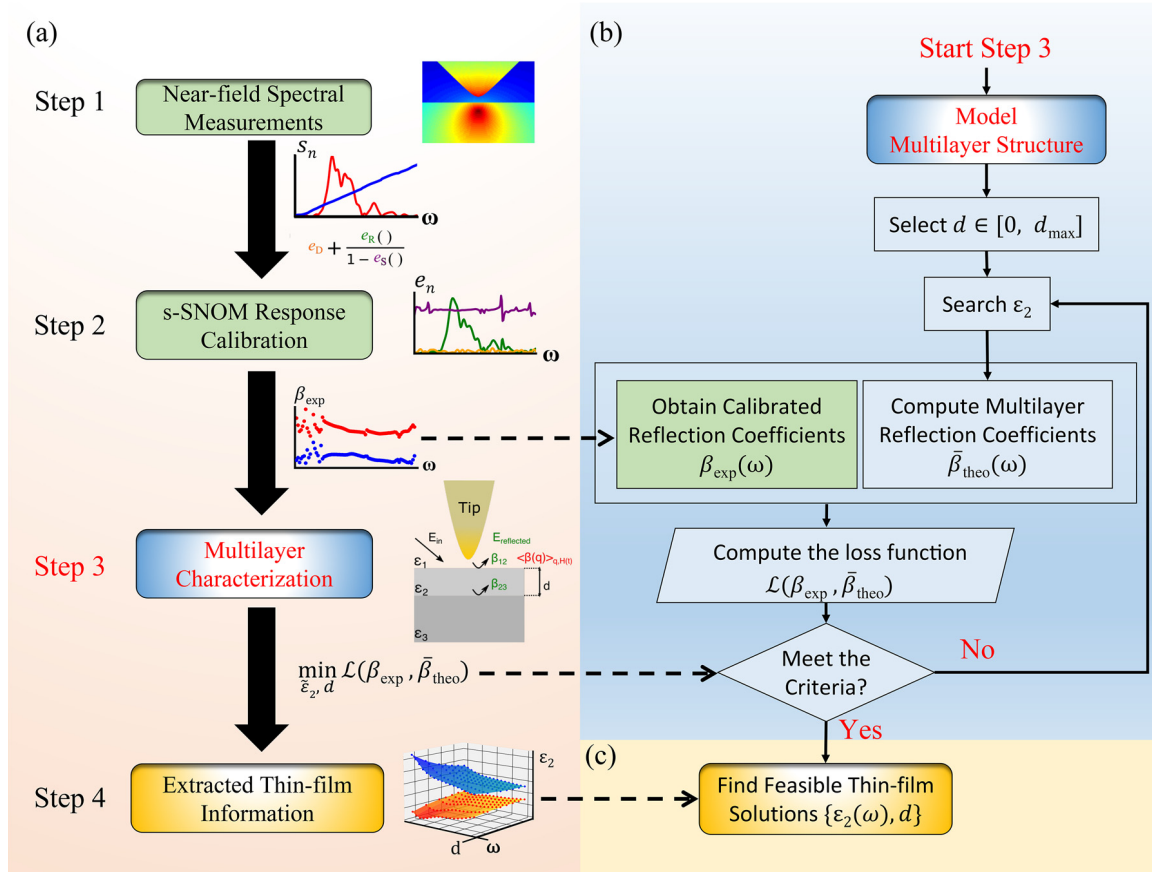

**Figure 3:** THz s-SNOM multilayer extraction procedure. (a) Flow diagram of the thin-film characterization employing THz s-SNOM with four steps. Details of multilayer characterization are displayed in (b, c).

knowledge of the superstrate (air: $\varepsilon_1$) and substrate (high-resistivity Si: $\varepsilon_3$).

- (4) obtain feasible solutions for the thin film after multilayer characterization and perform further analysis.

Here, we consider a multilayer structure with unknown thin-film thickness ($d$) and permittivity ($\varepsilon_2$) under the quasi-static limit with the in-plane momentum ($q$). Employing our previous vector-nature calibration method, we are able to calibrate the system responses with three error terms ($e_D$, $e_R$, $e_S$) and further recover the calibrated reflection coefficient as shown in step 1 and 2 in Figure 3(a).

The flow chart in Figure 3(b) describes the procedure to further extract the thin-film thickness and complex permittivity from the calibrated reflection coefficient ($\beta_{\text{exp}}$). To solve this inverse problem, first, a thin-film thickness ($d$) is selected. Then, an initial guess of complex permittivity ($\varepsilon_2$) is used to compute the multilayer reflection coefficient ($\beta_{\text{theo}}$). To find the feasible solutions, the deviation between the measured and the computed reflection coefficient, $\mathcal{L}(\beta_{\text{exp}}, \bar{\beta}_{\text{theo}})$, is minimized. An optimization problem is solved subject to the knowledge of a typical material: (1) the material is lossy and (2) not actively emitting energy to the environment, as formulated below:

$$\min_{\tilde{\varepsilon}_2, d} \mathcal{L}\left(\beta_{\text{exp}}, \bar{\beta}_{\text{theo}}\right)$$
$$\text{s.t. } \text{Re}\{\tilde{\varepsilon}_2\} \in \mathbb{R}$$
$$\text{Im}\{\tilde{\varepsilon}_2\} \geq 0$$
$$\text{Re}\left\{\sqrt{\tilde{\varepsilon}_2}\right\} \geq 1$$

This optimization problem is then solved repeatedly for each thickness in the predefined boundary for the thin-film thickness, $d \in [0, d_{\max}]$. In this study, we choose 100 nm as the thin-film upper limit. Finally, the optimal ($\varepsilon_2$, $d$) pairs are returned as the feasible solutions of the thin film for the multilayer structure. For each monochromatic response, an optimal ($\varepsilon_2$, $d$) pair exists, including both the real and imaginary parts of the permittivity. Therefore, the final extracted results for a thin film would be two parameter planes as shown in Figure 3(c).

The power of this method is demonstrated by the extraction of the permittivity of the thin film layer for the as-prepared silicon device discussed above from multiple harmonics ($S_1 - S_5$) of the near-field scattering signals. Figure 4(a) shows the extracted THz permittivity (real: solid lines, imaginary: dashed lines) between 0.6 and 1 THz (for results above 1 THz, see Figure S4). For the as-prepared sample, the real part of the thin-film permittivity is negative, indicating it holds an opposite sign with the substrate (high-resistivity silicon). This sign change corroborates with the dielectric conditions needed to support surface plasmon polaritons. We can then estimate the fabrication-induced doped layer thickness around 9–13 nm from Figure 4(c) according to the retrieved thickness from multiple orders ($S_1$–$S_5$) of harmonic signals. As the effective probing depth decreases with the increasing of harmonic orders, the intersection of the retrieved thickness from several higher-order harmonics further confirms the existence of a doped thin film on the high-resistivity silicon substrate.

To quantitatively describe the doped surface, in Figure 4(c), we compute the equivalent conductivity ($\sigma$) from the imaginary part of the permittivity, $\left(\sigma = \text{Im}\{\tilde{\varepsilon}_2\}\omega\varepsilon_0\right)$ where $\varepsilon_0$ is vacuum permittivity [34]. To better estimate the dopant concentration, we compute the uncertainty of the low-frequency THz responses (0.6–0.7 THz), as close to direct current conductivities, in our probing frequencies for extracted thin-film thickness. In addition, the thickness-dependent conductivity curve matches a typical exponential decay profile, $A \cdot \text{Exp}(-\alpha d)$, where $A$ is the initial amplitude and $\alpha$ is the decay constant.

To further validate the retrieved information of the thin film, the dispersion curve is reconstructed from the experimental data of the imaginary part of the p-polarized reflection coefficient at real-valued $q$. A representative thickness

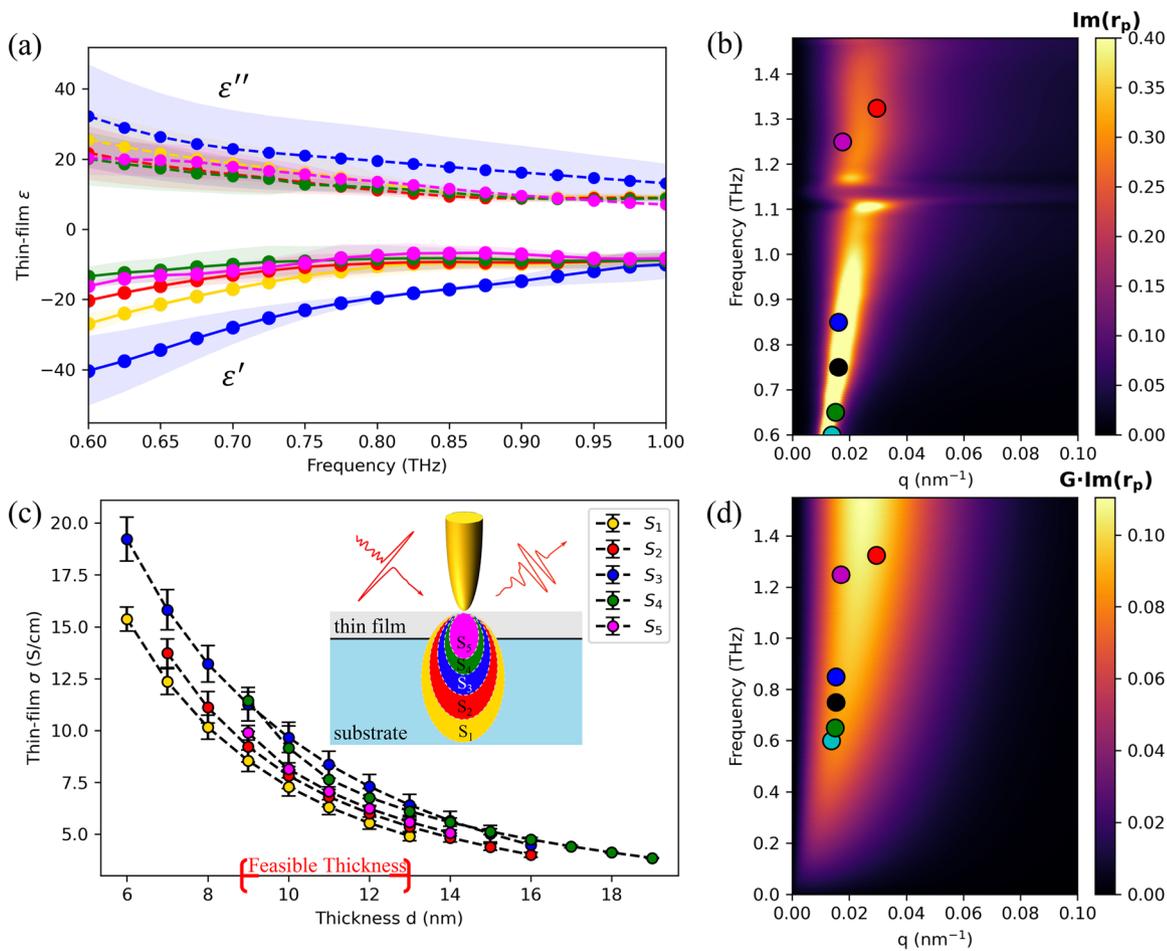

**Figure 4:** THz s-SNOM multilayer extraction results. (a) Extracted thin-film complex permittivity for the as-prepared sample with both the real (solid lines) and imaginary (dashed lines) parts. (c) Feasible solutions of thin film conductivity from near-field scattering signals demodulated at multiple harmonics ($S_1$–$S_5$) of the tip oscillation frequencies. Colours of the markers in panel (a) and (c) indicate harmonic orders and are consistent. Error bars are calculated from the uncertainty of the extracted thicknesses and conductivity in the selected probing frequency regions. The dispersion relation calculated by using (b) extracted thin-film thickness (13 nm) and permittivity from third-order harmonic signal ($S_3$) and by using (d) Drude model considering the multilayer nanostructure considered in Figure 2(c). The circles are in-plane momenta extracted from hyperspectral measurements at the corresponding THz frequencies.

(13 nm) and corresponding complex permittivity (extracted from $S_3$) are used to compute the curve in Figure 4(b). The feasible thin-film thickness range is determined by looking for solutions where the solutions converge and coexist across five harmonics as shown in Figure 4(c). Comparing with the theoretical dispersion curve in Figure 4(d), we notice that the retrieved in-plane momenta (circles) of near-field fringes from hyperspectral measurements matches well with the retrieved dispersion curves from s-SNOM measurements. For more retrieved information from different harmonic signals and thin-film thickness, see Figure S5. We also validate on a known multilayer nanostructure with $SiO_2$ on Si (TGQ1, TipsNano Co, Estonia). The extracted values of $SiO_2$ (see Figure S7) agree with the literature and sample specification [35]. We anticipate an increase of extraction accuracy on multilayer nanostructures across multiple-order harmonic signals in the future with advancements in signal-to-noise ratio on THz s-SNOM. Cases for nano-FTIR and s-SNOM measurements on other spectral ranges are subject to the further study.

## 2.4 Device performances at low temperature

Motivated by the extracted conductive thin-film information, we move further to selectively etch the silicon channel of the quantum device to remove the fabrication-induced surface doping. Low-temperature (∼50 mK) transmission measurements, as shown in Figure 5, are performed to determine quality factors of the fabricated resonators with varying degrees of etching. We find three dips around 5 GHz, 6 GHz and 7 GHz, which represent the response of those resonators. These three dips are then fit to obtain resonance frequencies and quality factors (internal ($Q_i$), external ($Q_c$), and loaded ($Q_l$) quality factors, see Methods).

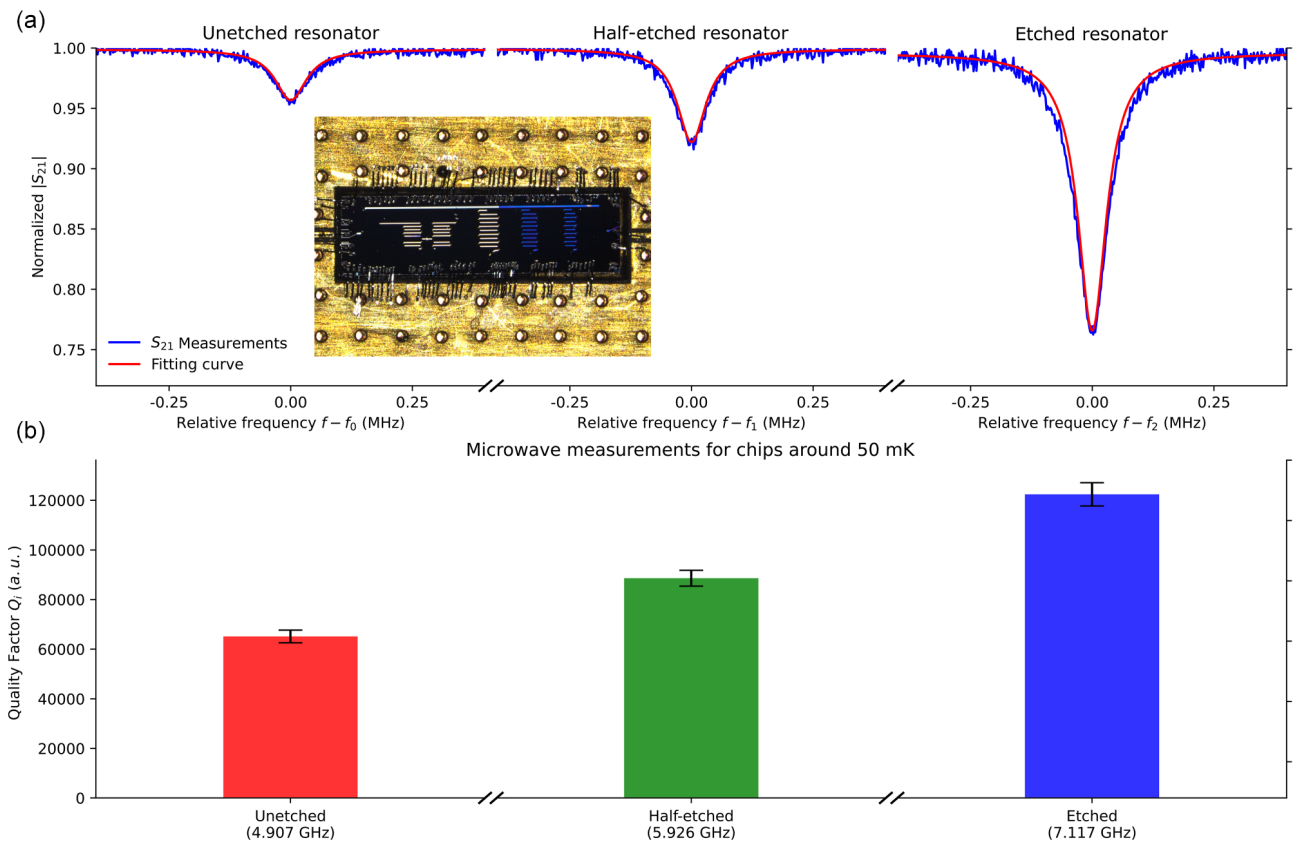

**Figure 5:** Low temperature quantum device performance. (a) Three dips indicate resonances at the characteristic frequencies ($f_0$ = 4.9 GHz, $f_1$ = 5.9 GHz, $f_2$ = 7.1 GHz) and are centred on zero for quality factor extraction. Inset: the mounted chip, where blue areas indicate etched resonators. (Half-etched resonator: half-coated to prevent etching, etched resonator: etching depth ∼200 nm. Blue colour under the optical microscope is due to the silicon trench). (b) Measured internal quality factor ($Q_i$) of fabricated resonators for quantifying losses due to two level fluctuations.

Table 1: Low-temperature microwave measurements for chips around 50 mK.

| Method | $f_r$ (GHz) | $Q_l$ (loaded) | $Q_i$ (internal) | $Q_c$ (coupling) |
| --- | --- | --- | --- | --- |
| Unetched | 4.907 | 62,132 ± 2384 | 64,863 ± 2555 | 1,475,694 ± 37,463 |
| Half-etched | 5.926 | 81,536 ± 2785 | 88,370 ± 3197 | 1,054,287 ± 21,855 |
| Etched | 7.117 | 93,911 ± 2887 | 122,147 ± 4692 | 406,243 ± 6063 |

The measured resonator quality factors are shown in Table. 1. Internal quality factor ($Q_i$), shown in Figure 5(b), is of interest since it quantifies internal losses caused by coupling to two level fluctuations, which may occur due to impurities in the substrate [2, 36]. An increase of internal quality factor is observed after the surface etching in the silicon channel, suggesting an effective removal of substrate dissipation loss dependent on the leakage conduction current [37].

# 3 Conclusions

In this article, we observe THz plasmon polaritons in prototypical superconducting quantum devices through THz s-SNOM hyperspectral nanospectroscopy and simulations. Our observations are supported by the existence of a 9–13 nm thick doped layer at the surface of the high-resistivity silicon substrate, thereby providing the necessary dielectric change of sign required to support polaritons. To verify the thickness and dielectric properties of the doped layer, we introduce a multilayer extraction procedure and quantitatively retrieve both the thickness and permittivity of the doped thin film layer using multiple s-SNOM harmonics. This multilayer extraction procedure does not require the prior knowledge of the permittivity model or dopants identity and is applicable to other devices structures. The convergence of multiple harmonics as well as the agreement between the hyperspectral experiment measurements and theoretical calculations corroborate the existence of nanoscale THz surface plasmons due to the doped thin film on top of the bulk silicon substrate. By selectively etching the surface of the silicon channel to mitigate the fabrication-induced surface doping, an increased quality factor is observed on the etched when compared to the unetched one. This further confirms the existence of fabrication-induced surface doping in our device. This study demonstrates that THz s-SNOM is an ideal tool for the non-destructive near-surface characterization of film thickness and dopant concentration and can be extended broadly to other nanodevices [38–40].

# 4 Materials and methods

## 4.1 Coplanar waveguide fabrication

High resistivity Si(100) (Topsil, floating zone grown, $\rho > 10k\Omega$ cm) was prepared by solvent cleaning, Piranha etching, and then etching in 5% buffer oxide etchant for 20 s to generate a hydrogen-passivated surface. The etched substrate was transferred to a Plassys MEB 550S electron beam evaporator and 100 nm of aluminium (1 nm/s by quartz microbalance) was deposited at room temperature. The sample is spin-coated with AZ1512-HS resist and patterned using direct-write lithography (Heidelberg Instruments µPG 101). Once the pattern is defined and developed (AZ726), aluminium is selectively etched from exposed regions using an etchant containing 21% deionized (DI) water, 73% $H_3PO_4$ 3% acetic acid, and 3% $HNO_3$ by volume. To selectively etch the silicon channel, reactive ion etching (RIE) is performed. The chip is sent to Oxford PlasmaPro80 to selectively etch the surface of exposed silicon substrate. AZ1512-HS resist is used to protect the underlying aluminium film from bombardment of ions. Flow of SF5 (20 ccm) and CHF3 (35 sccm) are introduced in the chamber with a pressure maintained at 10 mTorr. Ionized fluorine atoms react with Si under RF electric field (82 W). An additional piece of cleaned silicon wafer is used to cover the target area of the chip from its top side. When the etching is complete, the remaining photoresist is removed by submerging for 2 min in 60-degree VLSI acetone, followed by a 15 s rinse in VLSI isopropanol. The wafer is then dried with nitrogen gas.

## 4.2 THz s-SNOM measurements

Scattering-type near-field optical microscope (Neaspec GmbH, Germany) with a broadband THz time-domain spectroscopy (0.6–1.6 THz) was employed to interrogate the resonator chip. Commercially available atomic force microscopy probes (25PtIr200B-H, Rocky Mountain Nanotechnology LLC, U.S.A.) are used in this study. A 10-ps THz scattering field is collected from each spatial position (step size = 20 nm) by sweeping the optical scanning delay line. Probe tapping amplitude for THz near-field hyperspectral measurements is ∼290 nm.

## 4.3 Momentum-dependent p-polarized reflection coefficient

The momentum-dependent multilayer reflection coefficient $\beta(q)$ is given by:

$$\beta(q) = \frac{\beta_{12} + \beta_{12}e^{-2qd}}{1 + \beta_{12}\beta_{23}e^{-2qd}},$$

where the normal component of the wave vector is momentum-dependent, $k_z = iq$, under the electrostatic limit. Since the scattering signal measured by the receiver is momentum-averaged sampled through multiple periods of probe oscillation, the near-field reflection coefficient is given by [41–43]:

$$\overline{\beta}_{\text{theo}} = \left\langle \frac{\int_0^\infty \beta(q) q e^{-2qz} \mathrm{d}q}{\int_0^\infty q e^{-2qz} \mathrm{d}q} \right\rangle_z,$$

where $qe^{-2qz}$ is the tip-coupling weight function, $z = H(t) + w_{\text{eff}}$ expresses the distance between the effective monopole of the probe to the top layer of a multilayer nanostructure probed by an s-SNOM, $H(t) = a(1 - \cos(\omega t))$ describes the tip temporal oscillation during the near-field scattering measurements, $\omega$ is the probe oscillation frequency, $a$ is the probe tapping amplitude, and $w_{\text{eff}}$ describes the distance between the effective monopole and the tip apex. (See Figure S6(e) and (f) for a visual description) Convergence of the results for all five harmonics are achieved for $w_{\text{eff}} = 4$ nm. Realistic probe tapping amplitude $a = 290$ nm in THz nanospectroscopy measurements is used.

### 4.4 Fringe fitting procedure

To retrieve the polariton momenta from fringes in Figure 2(b), we use a Gaussian-type function as a fitting function:

$$S(x) = \left( \frac{A}{\sqrt{x - x_0}} e^{i(q' + iq'')(x - x_0)} \right) + B,$$

where $A$, $B$, $q'$ and $q''$ are the fitting parameters. In the fitting procedure, $x_0$ are kept as the constant for all fringes to determine the starting position for the fitting procedure. To avoid the edge artefact and considering the tip radius estimated from the edge-knife test (∼65 nm, see Extended Data, Figure S2), we choose to fit the fringe data starting ∼200 nm away (spatial position = ∼700 nm in Figure S3(d)) from the aluminium–silicon interface (spatial position = ∼500 nm in Figure S3(d)) till to the far end in the Si channel.

### 4.5 s-SNOM system response calibration

To reconstruct the reflectivity from the scattering amplitude, we perform a calibration using a vector method as described by Guo et al. [33] The calibration considers both, the intrinsic errors introduced via the SNOM itself and extrinsic errors from the environment (such as loss due to humidity or shifts in ambient temperature). It also acts as a converter taking the scattering near field signal and transforming it into the equivalent far-field reflectivity. The s-SNOM system response is calibrated using reference signals from a gold mirror, high-resistivity Si and air. These three standards allow us to extract the spatial-dependent complex THz reflectivity from hyperspectral data (see Extended Data, Figure S3). This calibration also allows direct extraction of complex pairs to describe materials under test (e.g., complex reflectivity or permittivity) without the requirement to fit the approach curve for calibrating tip-sample responses in the case of probe geometries are complex, which may be described by two probe geometry response factors in finite dipole model (see $f_{0,1}$ in Figure S6(d)) for mid-infrared cases.

### 4.6 Low-temperature characterization of quantum devices

For the low-temperature characterization of these devices, we packaged and mounted this chip onto the mixing chamber of a dilution refrigerator. Then the mixing chamber was cooled down to around 50 mK and the $S_{21}$ parameters were measured. We observed three dips around 5 GHz, 6 GHz and 7 GHz, representing the characteristic resonance frequencies of the three difference resonators. The microwave source was attenuated such that the average photon number circulating in the resonators was in the low-power regime (<1000 circulating photons). The three resonances were then fitted to obtain resonance frequencies and internal ($Q_i$), external ($Q_c$) and loaded ($Q_l$) quality factors using algorithms described in Probst et al.

**Acknowledgment:** The authors acknowledge the Traditional Owners and their custodianship of the lands on which UQ operates. We pay our respects to their Ancestors and their descendants, who continue cultural and spiritual connections to Country. The authors acknowledge the facilities, and the scientific and technical assistance, of the Microscopy Australia Facility at the Centre for Microscopy and Microanalysis, The University of Queensland. This work was performed in part at the Queensland node of the Australian National Fabrication Facility. A company established under the National Collaborative Research Infrastructure Strategy to provide nano and microfabrication facilities for Australia's researchers. XG acknowledges Prof. Rainer Hillenbrand for insightful discussions and suggestions provided in Munich.

**Author contributions:** XG and PJ initially drafted the manuscript. XG, KB, PJ and ADR designed the s-SNOM experiments. XG did s-SNOM experiments and analyzed the data. XH, ZD and CCC fabricated the quantum devices. CCC performed low-temperature measurements of devices. XG, BCD, KB, ZD, CCC, PJ and ADR revised the manuscript. AF, PJ and ADR supervised the project. All authors discussed the results, implications and contributed to the manuscript at various stages.

**Research funding:** Financial support was provided by the Australian Research Council (DP210103342) and the ARC Centre of Excellence for Engineered Quantum Systems (EQUS, CE170100009).

**Conflict of interest statement:** The authors declare no competing interests.

**Data availability:** The data generated and analyzed during this study are available from the corresponding author upon reasonable request.

# Supplementary Material: Terahertz nanospectroscopy of plasmon polaritons for the evaluation of doping in quantum devices

Xiao Guo[1], Xin He[2,3], Zachary Degnan[2,3], Chun-Ching Chiu[2,3], Bogdan C. Donose[1], Karl Bertling[1], Arkady Fedorov[2,3], Aleksandar D. Rakic[1*] and Peter Jacobson[2*]

[1*]School of Information Technology and Electrical Engineering, The University of Queensland, St Lucia, Brisbane, 4072, Queensland, Australia.
[2*]School of Mathematics and Physics, The University of Queensland, St Lucia, Brisbane, 4072, Queensland, Australia.
[3]ARC Centre of Excellence for Engineered Quantum Systems, St Lucia, Brisbane, 4072, Queensland, Australia.

*Corresponding author(s) E-mail(s): a.rakic@uq.edu.au; p.jacobson@uq.edu.au;
Contributing authors: xiao.guo@uq.edu.au; x.he@uq.edu.au; z.degnan@uq.edu.au; chunching.chiu@uq.edu.au; b.donose@uq.edu.au; k.bertling@uq.edu.au; a.fedorov@uq.edu.au; a.rakic@uq.edu.au; p.jacobson@uq.edu.au;

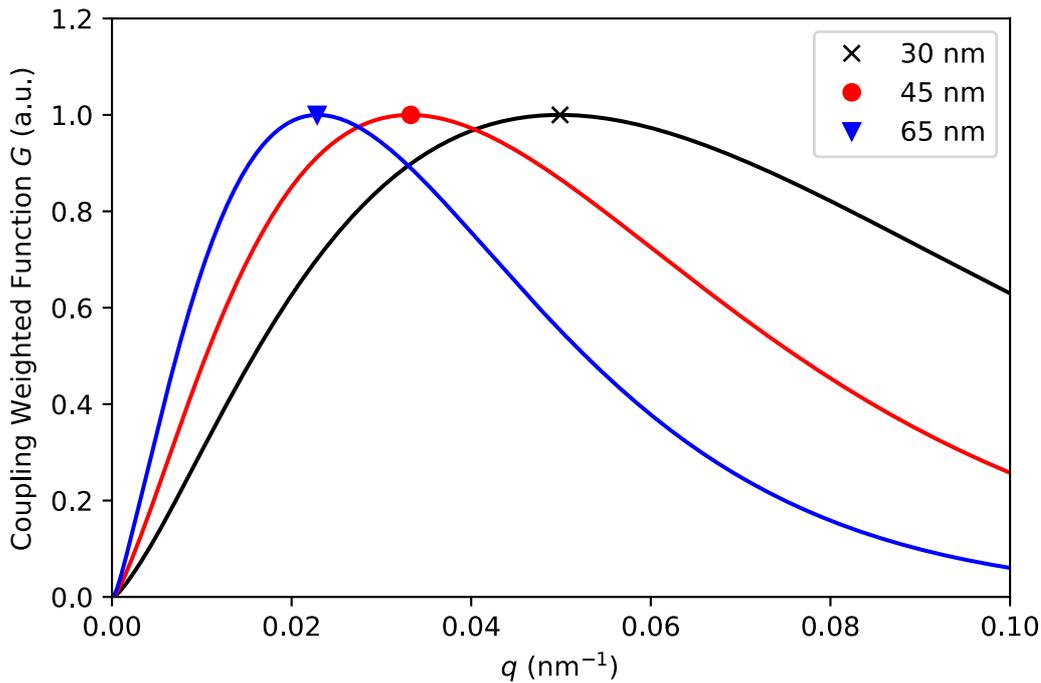

Fig. S1: a typical tip near-field coupling function, $< q^2 \exp(-2q(H(t)+b)) >_t$, for s-SNOM tip-sample interactions. q is in-plane momenta, $H(t) = A(1-\cos(\Omega t))$ is the probe tapping motion during the s-SNOM measurements, b is the distance between effective dipole and the tip apex, A is the probe tapping amplitude, $\Omega$ is the probe oscillation fundamental frequency, and t is time. We plot typical examples for tip radius to be 30 nm, 45 nm, and 65 nm.

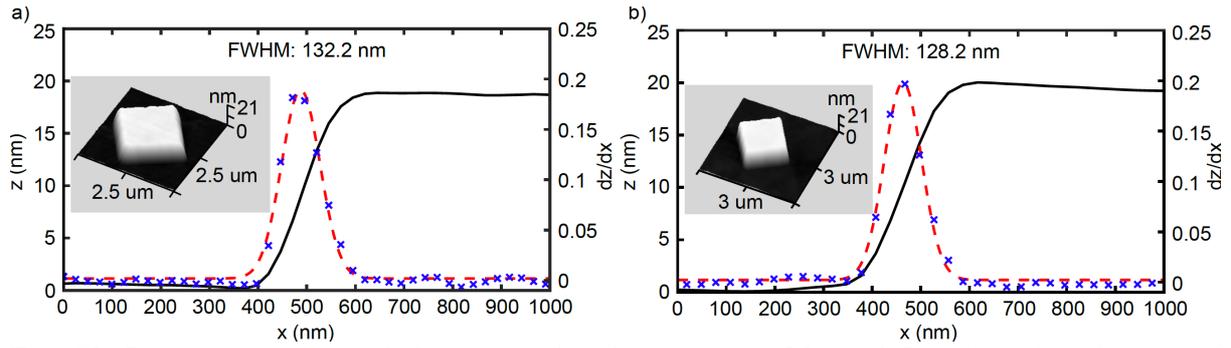

Fig. S2: Line scan across a SiO$_2$ square (height variation: 20 +/- 1.5 nm) on Si substrate of commercial AFM calibration sample (TGQ1, TipsNano Co, Estonia) with different probes (25PtIr200B-H, Rocky Mountain Nanotechnology). Insets are the corresponding topography scans. The left axes show the scanning height (z), and the right axes show the spatial gradient of height (dz/dx). A Gaussian kernel is used for fitting to extract the corresponding full-width half maximum (FWHM) to obtain the information of tip radius and spatial resolution.

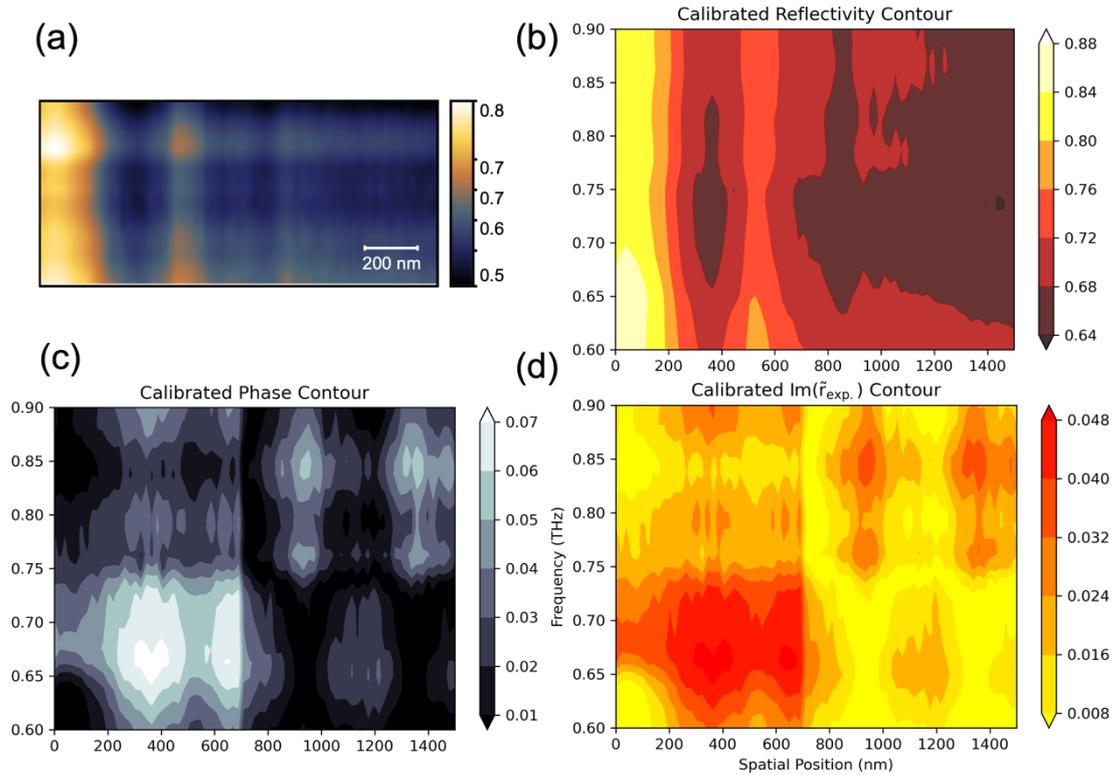

Fig. S3: Calibrated high-resolution (20nm/pixel) hyperspectral data, including **(a)** the amplitude of calibrated reflectivity from hyperspectral data up to 1 THz (with the horizontal axis identical to other three panels), **(b)** the contour of the calibrated reflectivity amplitude, **(c)** the contour of the calibrated reflectivity phase, and **(d)** the contour of the imaginary part of the complex calibrated reflectivity.

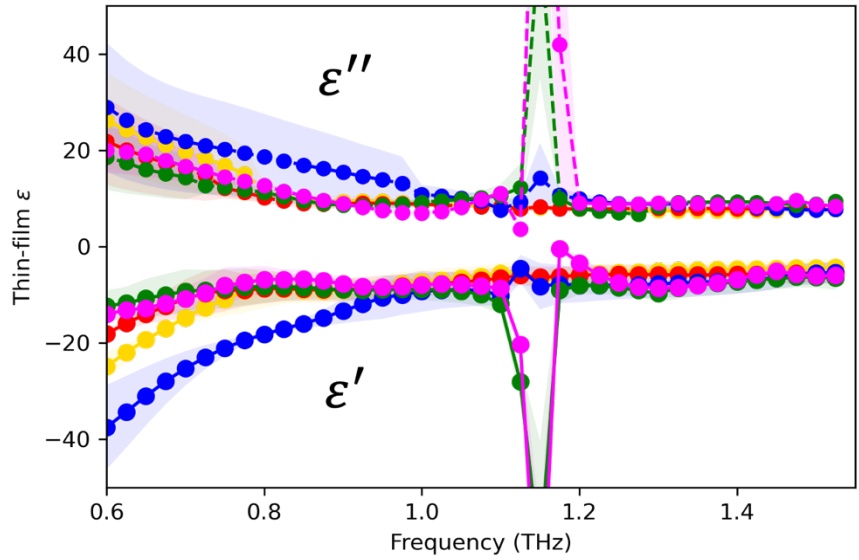

Figure. S4 Extracted complex permittivity of the as-prepared silicon device from 0.6 to 1.5 THz. The peaks around 1.1 THz are due to the existence of strong water vapour line.

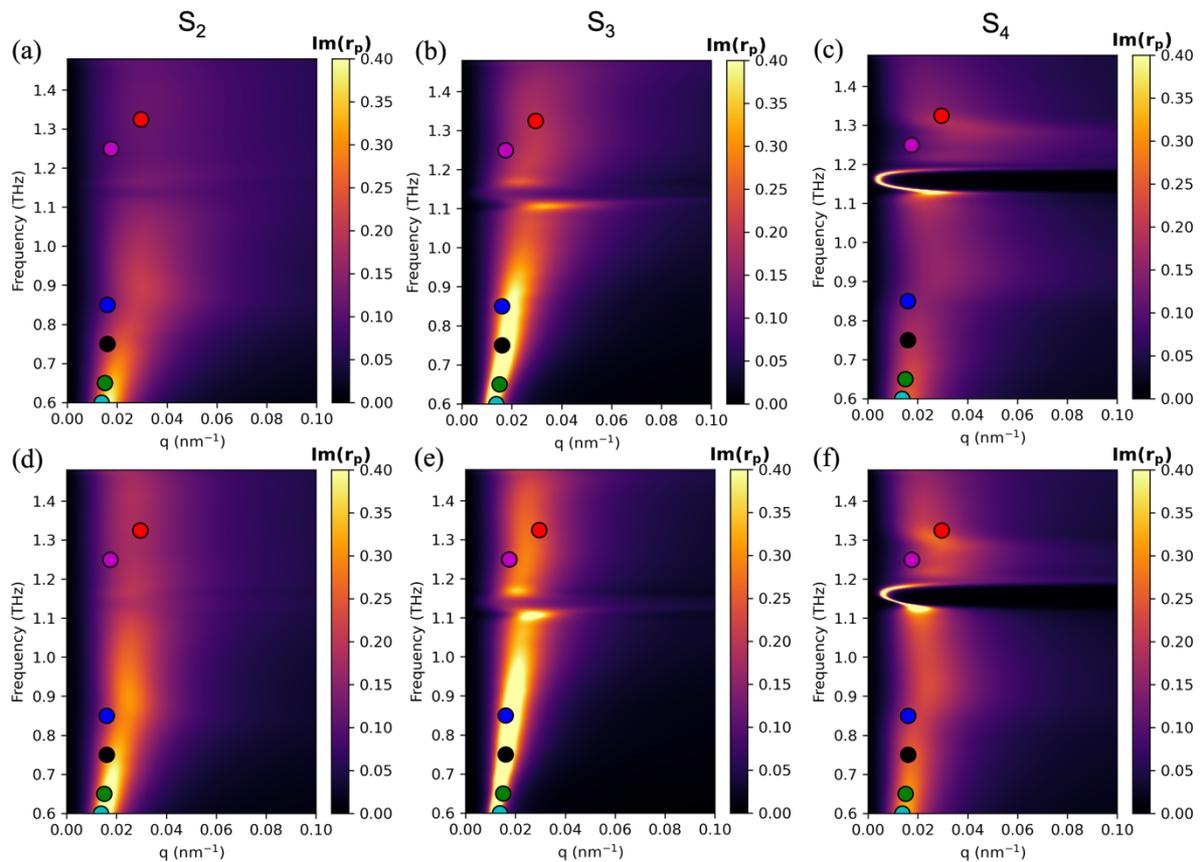

Figure. S5 The measured dispersion curve computed by using the retrieved thin-film thickness (9 nm (a, b, c) and 13 nm (d, e, f)) and complex permittivity from the second ($S_2$), third ($S_3$), and fourth ($S_4$) harmonic signals. The circles are extracted in-plane momenta from hyperspectral measurements at the corresponding THz frequencies.

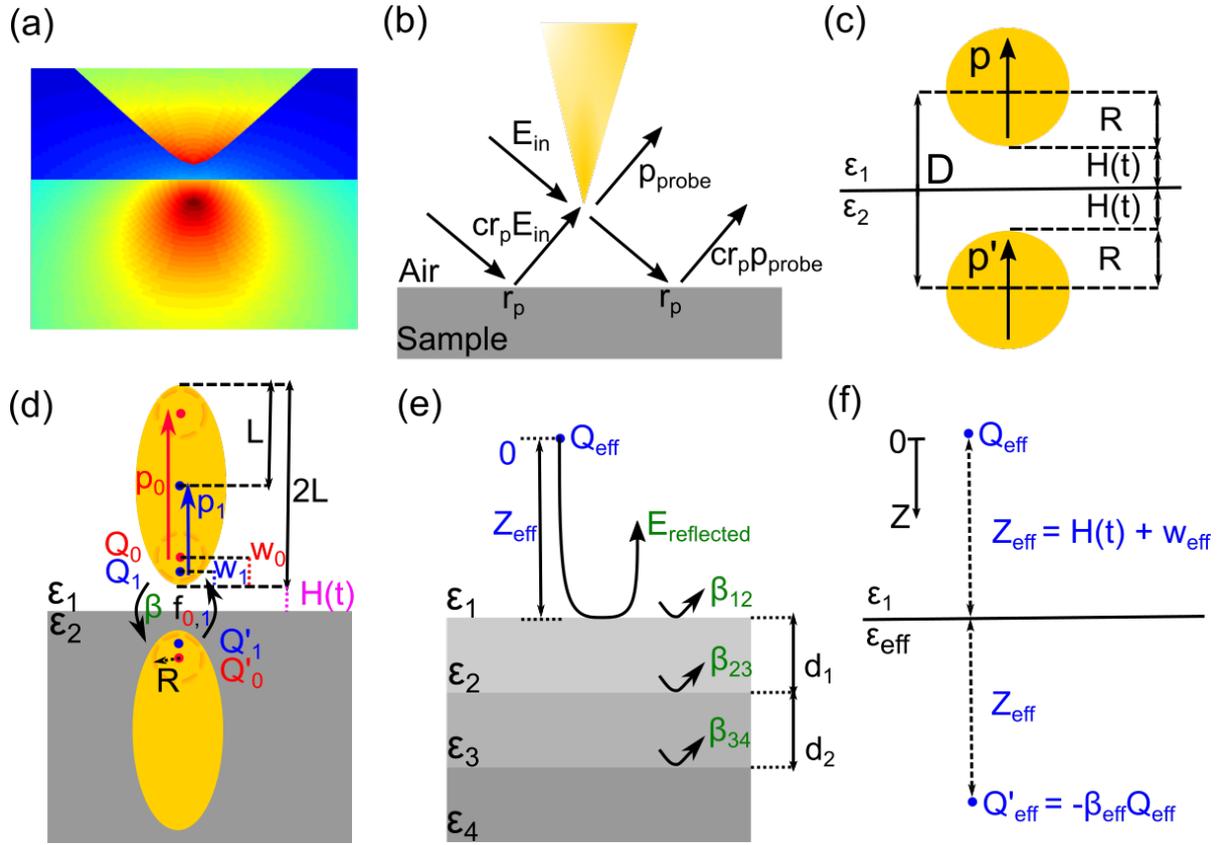

Figure. S6 Dipole effect in s-SNOM: (a) Simulation of the tip-sample's interaction between a 45-nm Au tip on Si substrate ~1 THz to show the s-SNOM near-field interaction. (b) Scattering process occurring in s-SNOM investigations. The s-SNOM tip is hovering at a height $H(t)$ above the sample with the permittivity $\epsilon$. The incident field $E_{in}$ is focused on the tip apex along with a contribution from the sample's surface reflection $cr_p E_{in}$ due to a wavelength-dependent beam focus (which is larger than the tip radius). The scattering signal is proportional to the overall momentum of the probe and consists of induced momentum $p_{probe}$ and $cr_p p_{probe}$ due to the near-field interaction and surface reflection. $r_p$ is transverse magnetic (TM or p) polarised Fresnel coefficient and $c = \exp(-i\Delta\phi)$ accounts for possible phase retardation between the direct radiation to/from the probe and reflection from the sample (which is usually regarded as 1 in s-SNOM analyses). (c) Point dipole model: the probe is approximated as a dielectric sphere with radius $R$ and effective momentum $p$ due to the near-field interaction. The investigated sample is at a distance $H(t)$ below the probe and mimicked as a mirror image with momentum $p'$ and the effective tip-probe distance is $D$. (d) Finite dipole model: the probe is approximated as an elongated spheroid (shank length $2L$, radius $R$) with the initial polarisation $p_0$ induced by the incident field $E_{in}$ and the near-field polarisation $p_1$ induced by the probe-sample coupling. It assumes only monopoles $Q_0$ and $Q_1$ near the sample surface participate in the near-field interaction. $W_{0,1}$ denotes the distance between monopole $Q_{0,1}$ and probe apex, $Q'_0$ and $Q'_1$ are the corresponding mirror images and $H(t)$ is the realistic tip-sample distance. (e, f) Multilayer treatments on near-field reflections: multiple reflections from the layered system (permittivity $\epsilon_i$ and thickness $d_i$) are approximated as an effective monopole $Q_{eff}$ at the distance $Z_{eff}$ above the surface. $w_{eff}$ is the effective distance between $Q_{eff}$ and the tip apex.

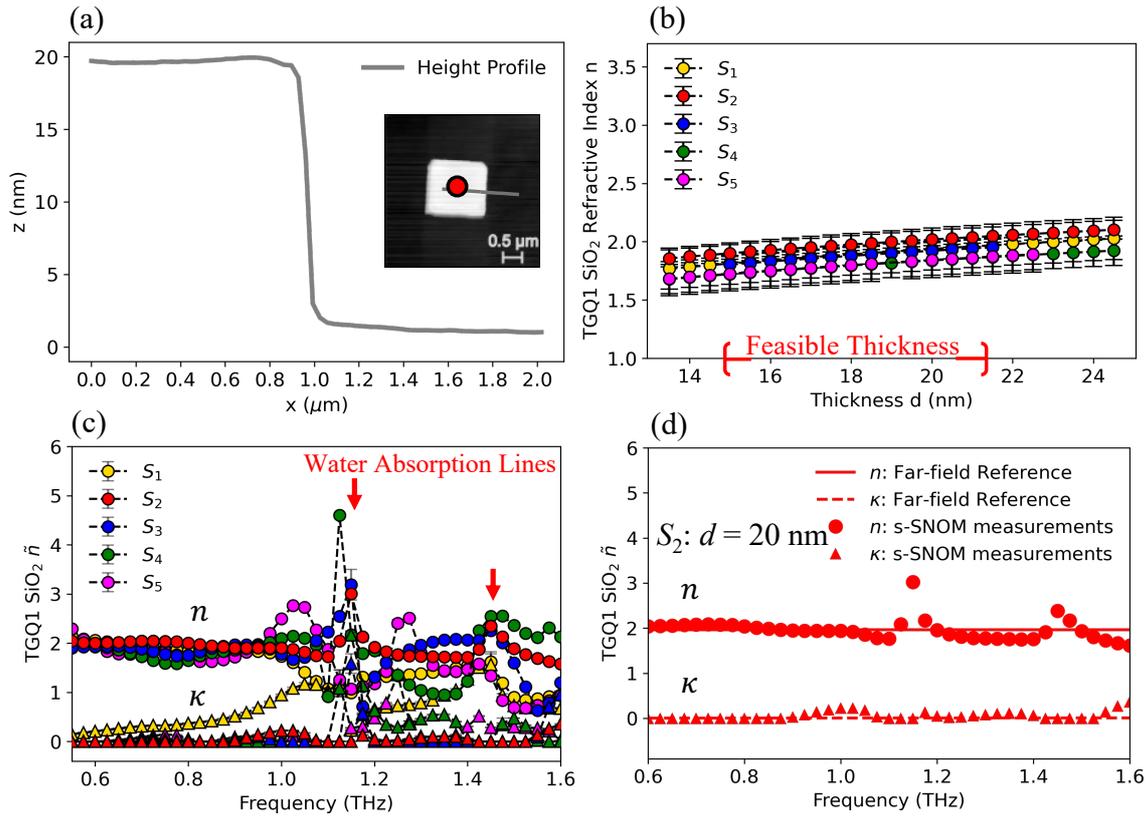

Fig. S7: THz s-SNOM multilayer extraction of complex-valued optical constant (optical constant: $n$, extinction coefficient: $\kappa$) and thickness ($d$) for a SiO$_2$ square (height variation: 20 +/- 1.5 nm) on a TGQ1 calibration sample: **(a)** a line profile scan across SiO$_2$ square and the substrate (see inset for the topography and positions to take the line profile); THz nanospectroscopy is taken at the centre of SiO$_2$ square (red circle in the inset); **(b)** Solutions of feasible thickness obtained from multiple harmonics ($S_1$ to $S_5$) of tip-scattered s-SNOM signals; **(c)** The extracted complex optical constants of SiO$_2$ square; Strong water absorption lines above 1 THz are indicated as red arrows; **(d)** Complex-valued optical constants from the 2$^{nd}$ harmonic signal ($S_2$) at 20 nm for SiO$_2$. Values from s-SNOM measurements are denoted as markers (circle: $n$, triangle: $\kappa$); Literature values of SiO$_2$ measured from far-field THz-TDS systems are shown as lines (solid: $n$, dash: $\kappa$).